\documentclass[twocolumn,prl,amsmath,amssymb,showpacs,superscriptaddress]{revtex4-1}
\usepackage{epsf}      
\usepackage{graphicx}
\usepackage{color}
\usepackage{soul}
\usepackage{gensymb}
\usepackage{sidecap}

\begin{document}

\title {Investigation of the structural, electronic, transport and magnetic properties of Co$_2$FeGa Heusler alloy nanoparticles}

\author{Priyanka Nehla}
\affiliation{Department of Physics, Indian Institute of Technology Delhi, Hauz Khas, New Delhi-110016, India}
\author{Clemens Ulrich}
\affiliation{School of Physics, The University of New South Wales, Kensington, 2052 NSW, Sydney, Australia}
\author{Rajendra S. Dhaka}
\email{rsdhaka@physics.iitd.ac.in}
\affiliation{Department of Physics, Indian Institute of Technology Delhi, Hauz Khas, New Delhi-110016, India}

\date{\today}
\begin{abstract}

We report the structural, transport, electronic, and magnetic properties of Co$_2$FeGa Heusler alloy nanoparticles. The Rietveld refinements of x-ray diffraction (XRD) data with the space group Fm$\bar {3}$m clearly demonstrates that the nanoparticles are of single phase. The particle size (D) decreases with increasing the SiO$_2$ concentration. The Bragg peak positions and the inter-planer spacing extracted from high-resolution transmission electron microscopy image and selected area electron diffraction are in well agreement with data obtained from XRD. The coercivity initially increases from 127~Oe to 208~Oe between D = 8.5~nm and 12.5~nm, following the D$^{-3/2}$ dependence and then decreases with further increasing D up to 21.5~nm with a D$^{-1}$ dependence, indicating the transition from single domain to multidomain regime. The effective magnetic anisotropic constant behaves similarly as coercivity, which confirms this transition. A complex scattering mechanisms have been fitted to explain the electronic transport properties of these nanoparticles. In addition we have studied core-level and valence band spectra using photoemission spectroscopy, which confirm the hybridization between $d$ states of Co/Fe. Further nanoparticle samples synthesized by co-precipitation method show higher saturation magnetization. The presence of Raman active modes can be associated with the high chemical ordering, which motivates for detailed temperature dependent structural investigation using synchrotron radiation and neutron sources.            
\end{abstract}
 
\maketitle

\section{\noindent ~Introduction}
 
Heusler alloys with the formula X$_2$YZ, where X, Y are transition metals and Z is a main group element, which posses the L2$_1$ structure have attracted great attention due to their rich potential for fundamental science and their wide range of practical applications in spintronics, for solar cells, thermoelectric applications, as shape memory materials, and topological semimetals  \cite{Felser16,Casper12,Jain17,Huanga16,Fujita01,ChangPRL17,Graf11}. In particular Co-based Heusler alloys are considered most important for future application due to their high Curie temperature (T$_{\rm C}$), well above room temperature and their ferromagnetic nature\cite{Felser07}. So far the Co$_2$FeSi alloy is known to posses the for highest T$_{\rm C}$ ($\sim$1100 K) with a magnetic moment of 6~$\mu_B/f.u.$  \cite{Wurmehl05,Gaier09}. In addition, many Co-based materials have been predicted to show half-metallicity i.e. 100\% spin polarization at the Fermi level \cite{Galanakis02}. However, the presence of some degree of structural disorder in these materials strongly reduces these properties while higher degree of ordering (L2$_1$/B2) stabelizes the spin polarization \cite{ElmersPRB03,MiuraPRB04}. For developing next generation of spin related and ferromagnetic shape memory devices, the understanding of the physical properties of Heusler alloys at the nanoscale is of vital importance and will play a decisive role \cite{Chris, WangJAP07, LiuJAP07}.  

In recent years, the Co-based Heusler alloys have been studied by reducing the dimensionality, which envisage the importance of size and interfaces in altering their physical properties for various applications \cite{Zayak08,WangJVST14}. 
However, the synthesis procedure of Heusler alloy nanoparticles and controlling the particle size have its own limitations such as the requirement of annealing at high temperatures under hydrogen environment \cite{Basit09,Wang10} and an appropriate annealing rate \cite{Kashi16}. Basit \textit{et al.}, synthesized and investigated ternary Co$_2$FeGa nanoparticles by simple chemical method \cite{Basit09} and it was determined that the short range ordering in these nanoparticles strongly influences their structural and magnetic properties \cite{Wang12,Wang17}. Recently, Wang \textit{et al.} studied the size dependent structural and magnetic properties of Co-Ni-Ga shape memory Heusler alloy nanoparticles, which report that the saturation magnetization as well as T$_{\rm C}$ increase while the coercivity decreases with increasing the particle size in the nanoscale range \cite{Wang17}. Note that the authors in refs.~\cite{Basit09, Wang12, Wang17} have used fumed silica powder as template for preparing nanoparticles. Alternatively, another simple and effective co-precipitation route for preparing Heusler alloy nanoparticles has been reported recently by  Yang {\it et al.} \cite{YangCPL17}. The authors have studied the effect of different pH values on the structural and magnetic properties, when preparing the Co-Fe-Al nanoparticles in a solution of water and NaOH. For a pH value of 7.0 small particle size with B2 ordering was observed, which did exhibit a high saturation magnetization and a low coercivity \cite{YangCPL17}. 

However, a detailed study of the magnetic domain structure, critical size, and magnetic anisotropy is still lacking in Co$_2$FeGa nanoparticles, and a systematic investigation is required for the various technological applications. In order to understand the relationship between particle size and magnetic properties it is necessary to control the particle size during the synthesis of nanoparticles. In this paper we have focused on the correlation of domain structure and particle size of Co$_2$FeGa nanoparticles. The nanoparticle samples have been synthesized by the sol-gel method with silica support, where the silica concentration plays a crucial role in controlling the particle size \cite{WangCM10}. We have studies how the particle size affects the magnetic properties and discuss the results in terms of the transition from single domain to multi-domain regime. We have also synthesized Co$_2$FeGa nanoparticles by a simple co-precipitation method and have investigated the effect of the pH value of the water-NaOH solution on their structural and magnetics properties. The analysis of resistivity data of CFG-25 sample indicates a complex scattering mechanism in the electron transport. Further, we characterize our samples using photoemission (XPS and UPS) and Raman spectroscopy techniques to understand electronic structure and lattice vibrations, respectively.   

\section{\noindent ~Experimental Details}

Co$_2$FeGa nanoparticles were synthesized by the sol-gel method, where 0.48 mmol of Fe(NO$_3$)$_3$.9H$_2$O, 0.49 mmol of CoCl$_2$.6H$_2$O and 0.32 mmol of Ga(NO$_3$)$_3$.xH$_2$O (x=8) were mixed in 50 ml methanol. All the precursors were used as purchased from a commercial supplier without any further purification. For decomposition and reduction of Co, Fe and Ga precursor the solution was sonicated for 5~min. Further, a known concentration of fumed silica (1000, 750, 500, 250, 100, 50 and 25~mg) was added to this solution and mixed in ultra-sonicator bath for 1h. After that when the resulting solution turned  yellowish, it was continuously stirred  overnight at room temperature to evaporate the solvent and obtain in the gel form. Then, this gel has been dried at 90$^0$C for 2hrs. In order to get fine homogeneous dense powder, this solid was gently grinded for few minutes and annealed at 800$^0$C for 5hrs in 10\% H$_2$ balanced Ar flow \cite{WangCM10,Priyanka18}. The final product was allowed to cool up to the room temperature and collected to study the structural and magnetic properties. We name the samples as CFG-1000, CFG-750, CFG-500, CFG-250, CFG-100, CFG-50 and CFG-25, which means the samples prepared in the presence of 1000~mg, 750~mg, 500~mg, 250~mg, 100~mg, 50~mg and 25~mg SiO$_2$ respectively.

An another series of Co$_2$FeGa nanoparticles samples have been synthesized by co-precipitation method as follow. Initially all the precursors in the proper stoichiometric form, i.e. CoCl$_2$.6H$_2$O (0.400g), Fe(NO$_3$)$_3$.9H$_2$O (0.404g) and Ga(NO$_3$).8H$_2$O (0.500g) have been dissolved in 30~ml de-ionized water and kept to stir for complete mixing of these precursors. Then, a solution of NaOH was added drop wise during the stirring and the pH value was adjusted accordingly. When all the metal ions are settled, the precipitate was obtained by filtering and washing this solution by de-ionized water. Finally, the precipitate was annealed at 800$^0$C for 2hrs in gas mixture of Ar with 10~\% H$_2$. In order to study the effect of pH on the physical properties of these nanoparticles, we varied the pH values from 5, 7, 9, to11 and named the samples as pH5, pH7, pH9, and pH11, respectively.

X-ray diffraction (XRD) measurements were performed using Cu K$\alpha$ ($\lambda$ = 1.5406~$\rm \AA$) radiation and the obtained data were analyzed by the Rietveld refinements using the FullProf package, where the background was fitted using a linear interpolation between the data points. For transmission electron microscopy (TEM) measurement the samples were transferred from methanol suspension to copper grids. The TEM and selected area electron diffraction (SAED) measurements were performed using the JEOL JEM-1400 Plus microscope with 120~KV accelerating voltage. Magnetic susceptibility and transport measurements were performed using a physical properties measurements system (PPMS EVERCOOL-II) from Quantum design, USA. A commercial photoemission spectrometer (ESCALAB250Xi from Thermo, located at the UNSW, Australia) with chamber pressure $<$ 5$\times$10$^{-10}$~mbar has been used to study the electronic structure of the CFG-25 sample. We used a monochromatic AlK$_{\alpha}$ x-ray source (h$\nu=$ 1486.6~eV) and a discharge He source (h$\nu=$ 21.2~eV) for performing x-ray photoemission spectroscopy (XPS) and ultraviolet photoemission spectroscopy (UPS) measurements, respectively. For Raman spectroscopy measurements, we used a Argon gas laser with excitation wavelength of 514~nm. The scattered light was analyzed using a Dilor-XY triple spectrometer (at UNSW) in backscattering geometry. In order to avoid laser heating of the metallic samples, the power of the incident laser beam was kept below 10~mW at the sample position.

\begin{figure}
\includegraphics[width=3.6in]{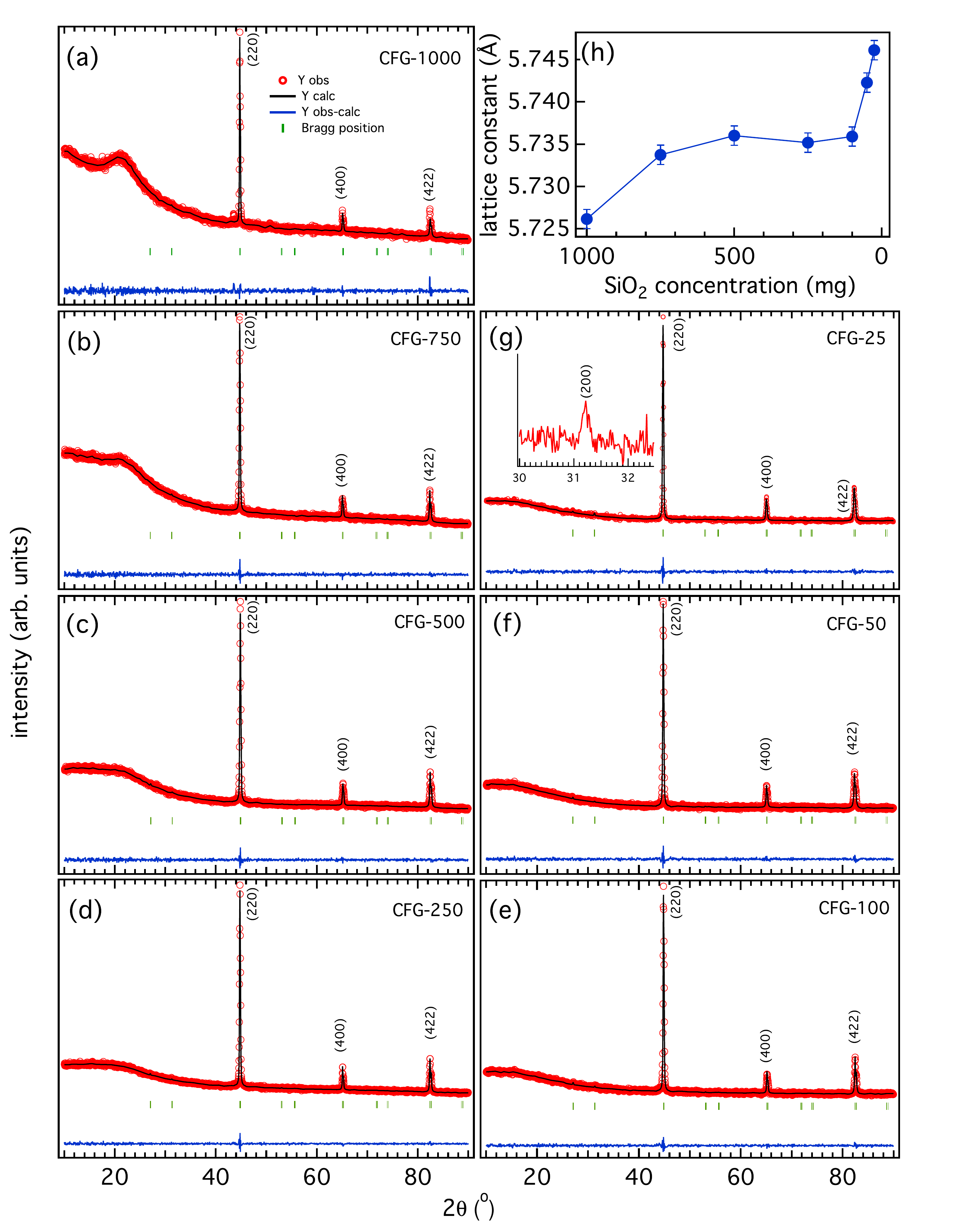}
\caption{(a-g) Powder XRD data (open circles) and Rietveld refinement (black line) of Co$_2$FeGa nanoparticles with different SiO$_2$ concentration along with the difference profile (blue line) and Bragg peak positions (vertical green bars). (h) Variation of lattice constant with SiO$_2$ concentration.}
\label{Fig1_FCG_allSiO2_new}
\end{figure}

\section{\noindent ~Results and Discussion}

\begin{table*}[ht]
\caption{Lattice constant ($a$) obtained by Rietveld refinements, saturation magnetizaton (M$_{\rm S}$), coercivity (H$_{\rm C}$) and the particle size (D) of Co$_2$FeGa nanoparticles for different  SiO$_2$ concentration.}
\vskip 0.2 cm
\centering
\begin{tabular}{c c c c c c c c} 
\hline\hline 
  & CFG-1000 & CFG-750 & CFG-500 & CFG-250 & CFG-100 & CFG-50 & CFG-25\\ [0.5ex] 
\hline 
$a$ ($\rm \AA$) & 5.7261(5) & 5.7337(9) & 5.7360(4) & 5.7351(8) & 5.7359(0) & 5.7422(6) & 5.7460(8) \\ 
M$_{\rm S}$ ($\mu_{\rm B}/f.u.$) & 0.6 & 0.55 & 0.95 & 1.6 & 2.9 & 2.75 & 3.5\\
$H_{\rm C}$ (Oe) & 127 & 177 & 208 & 186 & 150 & 100 & 68\\
D (nm) & 8.5 $\pm$ 0.3 & 10.5 $\pm$ 0.2 & 12.5 $\pm$ 0.2 & 16 $\pm$ 0.2 & 19 $\pm$ 0.2 & 19.5 $\pm$ 0.4 & 21.5 $\pm$ 0.5\\
\hline 
\end{tabular}
\label{table1} 
\end{table*}

Figure~\ref{Fig1_FCG_allSiO2_new} compares the XRD patterns of Co$_2$FeGa nanoparticles, prepared with different concentration of SiO$_2$. The main Bragg peaks of single phase Co$_2$FeGa, i.e. (220), (400), and (422) are observed at 2$\theta$= 44.5, 65 and 82$^o$, respectively. We have performed the Rietveld refinement of powder XRD data using the space group Fm$\bar{3}$m (no. 225). The crystal structure is stable for all the nanoparticles with different SiO$_2$ concentration. The lattice constant shows an increasing trend with decreasing SiO$_2$ concentration, as shown in Fig. \ref{Fig1_FCG_allSiO2_new}(h) and summarized in Table \ref{table1}. The presence of the (200) Bragg peak for the sample CFG-25, as shown in the inset of Fig.~1(g) and the absence of the (111) peak indicates the B2 disorder in this nanoparticle sample. It should be noted that a hump is observed in the XRD data at about 2$\theta$ $\approx$ 20$^0$. This hump can be attributed to SiO$_2$ and is found to be decreasing with decreasing the SiO$_2$ amount, as seen in Figs.~\ref{Fig1_FCG_allSiO2_new}(a-g).

\begin{figure}
\includegraphics[width=3.35in]{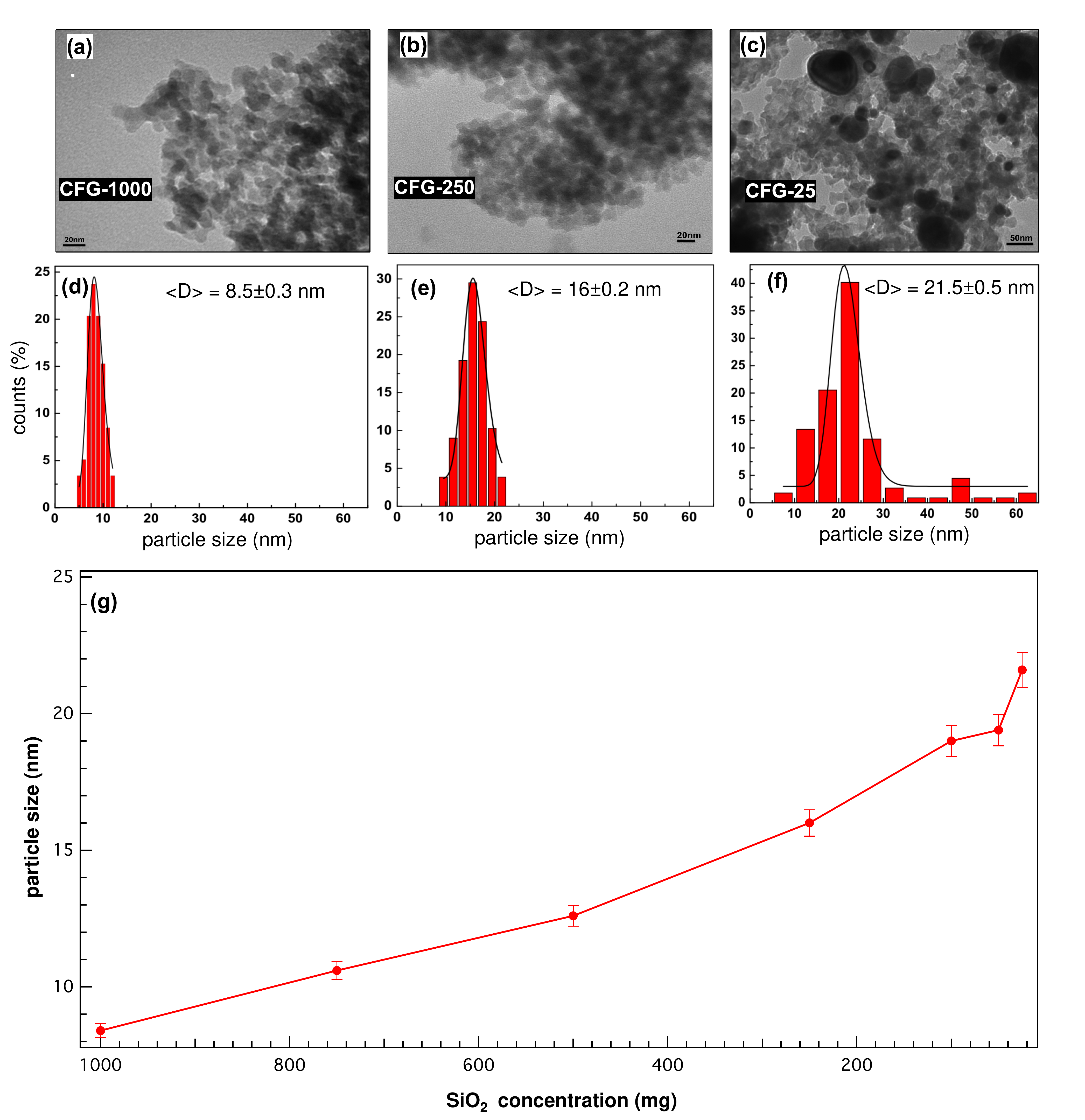}
\caption{TEM images (a--c) and fitted particle size distribution histograms (d--f) of the CFG-1000 (a, d) , CFG-250 (b, e), and CFG-25 (c, f) nanoparticles, respectively and (g) resulting particle size variation with SiO$_2$ concentration.}
\label{Fig2_all_TEM1}
\end{figure}

\begin{figure}
\includegraphics[width=3.5in]{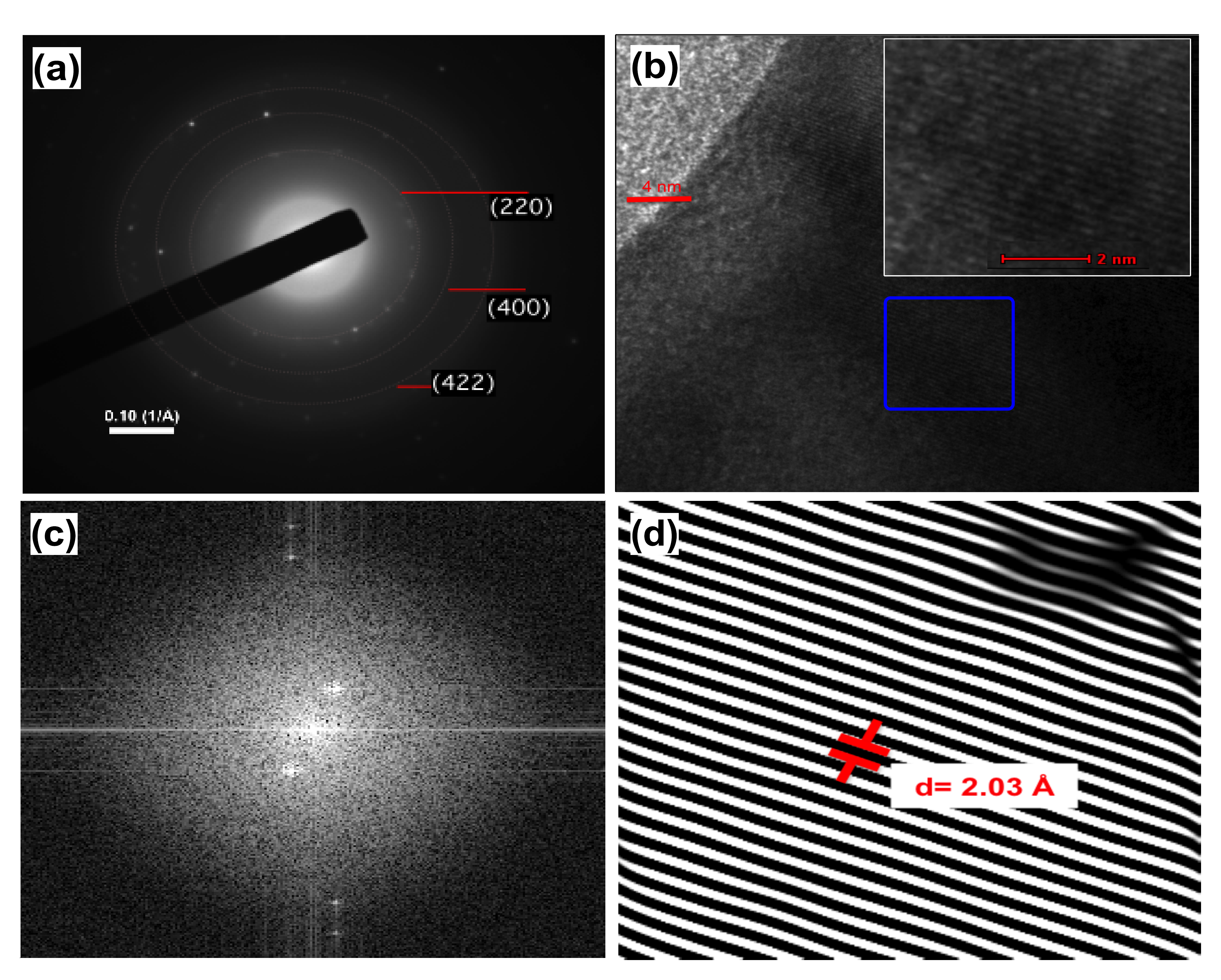}
\caption{(a) The SAED pattern and (b) high-resolution TEM image (inset is the zoomed view), (c) FFT, (d) inverse FFT pattern of selected area from (b) of CFG-25 sample.}
\label{Fig3_CFG25_SAD_HRTEM_FFT_IFFT}
\end{figure}

The particle size ($\textless {\rm D} \textgreater$) distributions of the Co$_2$FeGa nanoparticles was studied using TEM. In Figs.~\ref{Fig2_all_TEM1}, we show the TEM images (a-c) and the extracted particle size distribution histograms (d-f) of CFG-1000, CFG-250 and CFG-25 nanoparticle samples, respectively. All TEM images show that the particles are roughly spherical in shape. We have plotted the particle size histogram by considering a large number of individual particles (around 100) using ImageJ software. The histograms have been fitted with lognormal fit (solid line), in which the peaks correspond to the average particle size. From the histogram it is clearly seen that the particle size and the distribution are increasing with decreasing the SiO$_2$ concentration, see Figs. \ref{Fig2_all_TEM1}(d--f). The variation of the particle size for all prepared Co$_2$FeGa nanoparticles is shown in Fig. \ref{Fig2_all_TEM1}(g). We observe an increase in the particle size with decreasing SiO$_2$ concentration from $\textless {\rm D} \textgreater$ = 8.5~nm to 21.5~nm for the CFG-1000 and CFG-25 nanoparticle samples, respectively. As next step the crystallinity of the nanoparticles has been studied by SAED [see Fig.~3(a)]. The diffraction peaks observed in SAED are consistent with XRD results, which confirms that the nanoparticles have a well defined crystalline single ordered phase. The high-resolution TEM image of the CFG-25 sample, as shown in Fig.~3(b), clearly have the expected fringes corresponding to (220) plane, as also evident in the inverse FFT [see Figs.~\ref{Fig3_CFG25_SAD_HRTEM_FFT_IFFT}(c, d)].

\begin{figure}
\includegraphics[width=3.55in]{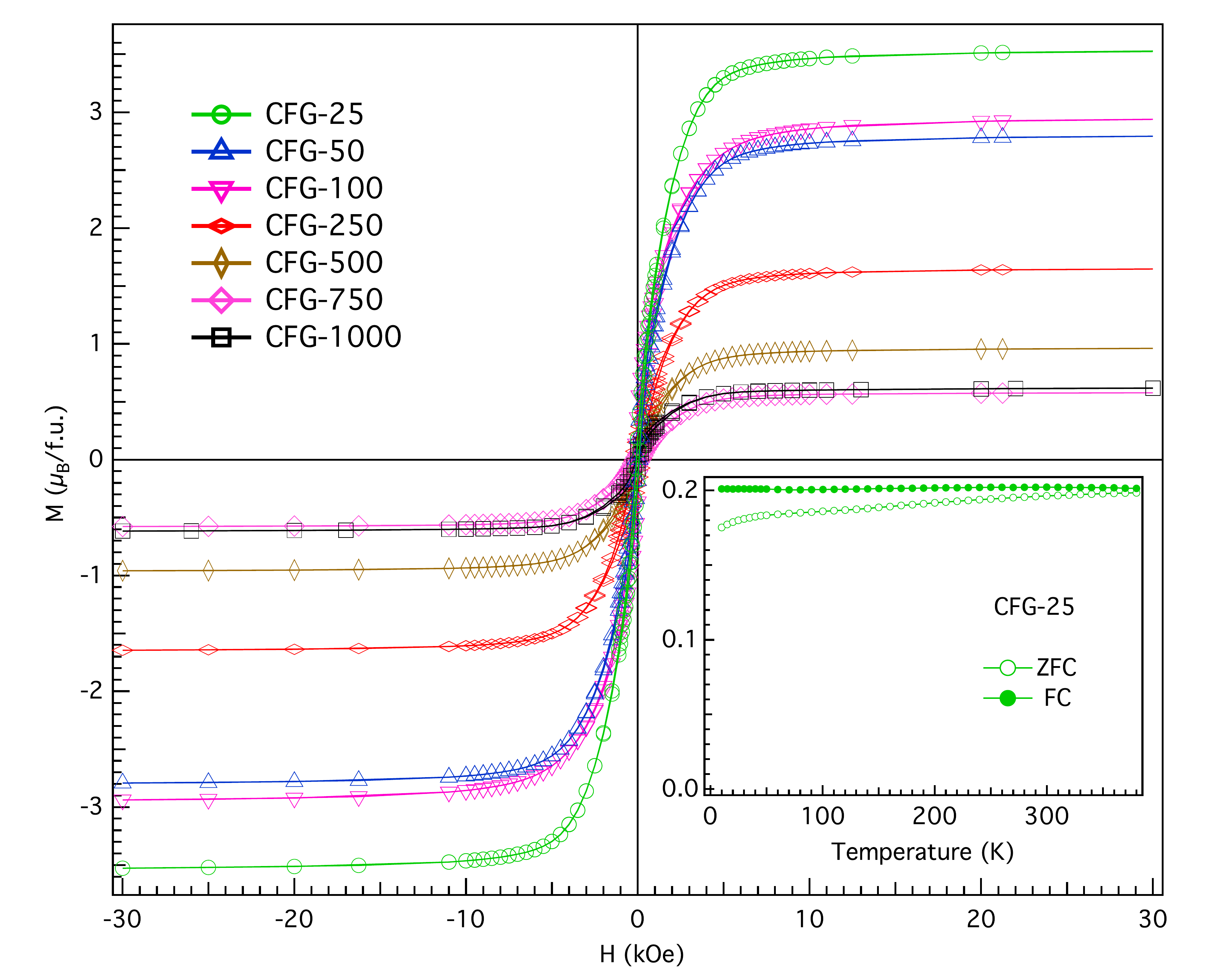}
\caption{The magnetization vs field curves of different Co$_2$FeGa nanoparticle samples measured at 300~K. The inset shows ZFC and FC curves of CFG-25 sample measured at 100~Oe.}
\label{Fig4_MH-MT_CFG25_new}
\end{figure}

In order to study the magnetic properties, we have measured the isothermal magnetization curves of Co$_2$FeGa nanoparticles, measured at room temperature (see Fig.~\ref{Fig4_MH-MT_CFG25_new}). Note that the half-metallic full Heusler alloys follow a Slater-Pauling behavior \cite{Kubler84}. Interestingly, the saturation magnetizaton (M$_{\rm S}$) increases, whereas the coercivity first increases and then decreases with increasing $\textless {\rm D} \textgreater$, as summarized in Table \ref{table1}. The highest M$_{\rm S}$ value is found for the sample CFG-25 and is found to be about 3.5 $\mu_{\rm B}/f.u.$, which is close to the expected value ($\approx$5 $\mu_{\rm B}/f.u.$) from Slater-Pauling rule. The difference can be attributed to the presence of disordering in the samples and/or different magnetic interactions in these nanoparticles. The inset in Fig. \ref{Fig4_MH-MT_CFG25_new} shows the changes in the magnetization with temperature in which both zero-field-cooled (ZFC) and field-cooled (FC) measurements have been taken at 100~Oe in the temperature range 10--380~K. In this temperature range we did not observe any transition. Since the Curie temperature (T$_C$) of bulk Co$_2$FeGa has been reported around 1100~K \cite{UmetsuPRB05}, higher temperature measurements are required to understand the magnetic transitions in these nanoparticles.

\begin{figure}
\includegraphics[width=3.45in]{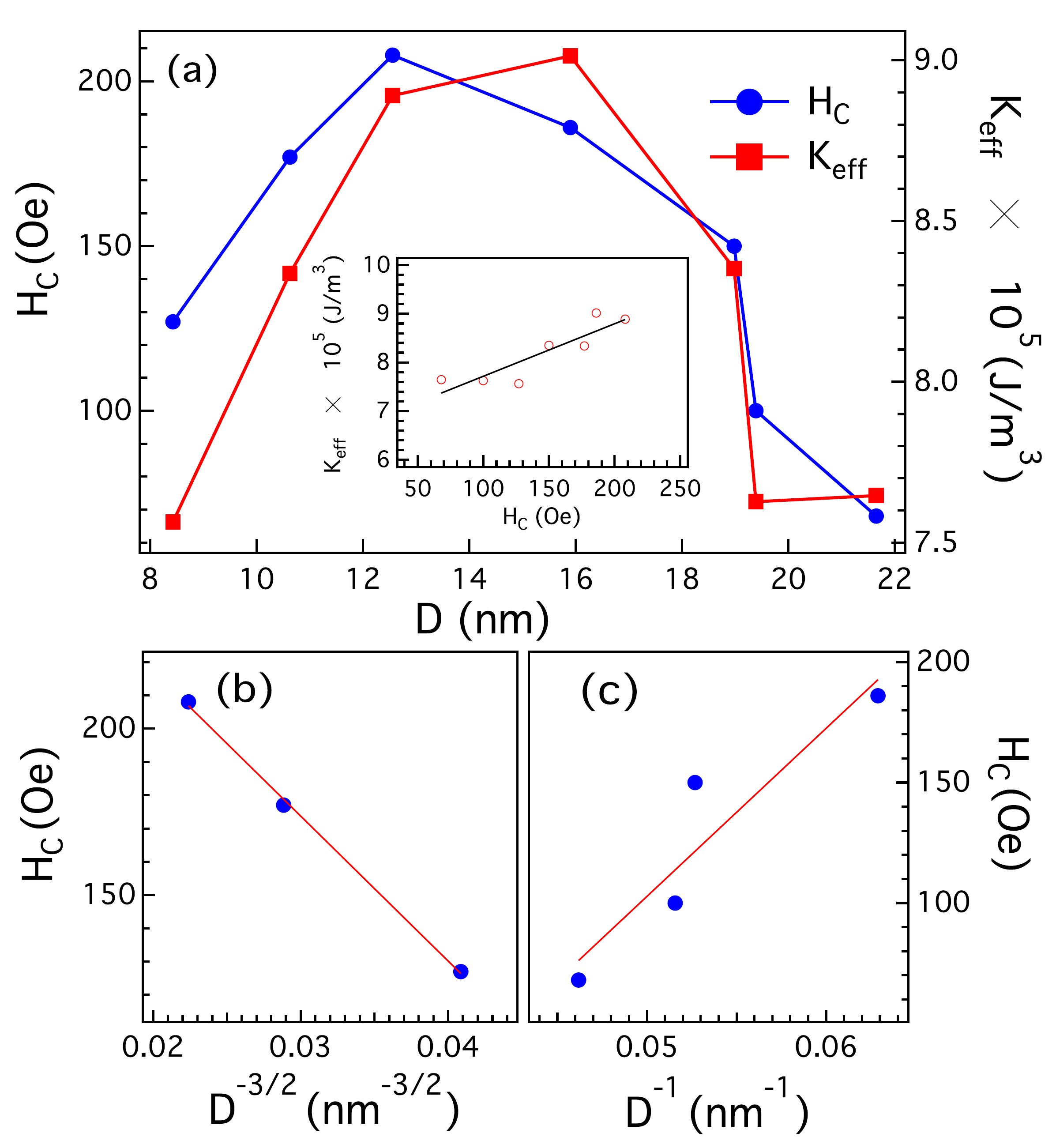}
\caption{(a) H$_{\rm C}$ and $K_{eff}$ variation with particle size (D). The inset shows the linear relation between $K_{eff}$ and H$_{\rm C}$. (b) H$_{\rm C}$ fitted with $D^{-3/2}$ in the 8.4~nm$ < $D$ < $12.6~nm range  and (c) fitted with $D^{-1}$ in the 16~nm$ < $D$ < $21.6~nm range.}
\label{Fig5_K_vs_Hc1}
\end{figure}

Recently, Li {\it et al.}, studied the correlation between the particle size and magnetic properties of Fe$_3$O$_4$ nanoparticles and discussed the results in terms of single and multidomain structures \cite{Li17}. In order to understand our results, here we have plotted, in Fig.\ref{Fig5_K_vs_Hc1}(a), the variation of coercivity ($H_{\rm C}$) with the particle size ($\textless {\rm D} \textgreater$). Interestingly, the H$_{\rm C}$ value increases and attains a maximum of 208~Oe at $\textless {\rm D} \textgreater$ = 12.5~nm. With further increase in the particle size, the H$_{\rm C}$ starts decreasing. This type of behavior of H$_{\rm C}$ has been reported for Fe$_3$O$_4$ nanoparticles \cite{Li17} and Ni-Zn ferrite \cite{Jiang11}. In order to analyse the magnetic behavior (Fig. \ref{Fig4_MH-MT_CFG25_new}) of the nanoparticles with the particle size in more detail, we have calculated the effective magnetic anisotropy constant, $K_{eff}$ using following relation \cite{Liu06}:
\begin{equation}
M(H)=M_S(1-\frac{0.0762 K_{eff}^2}{{M_S}^2H^2}-...)+\chi_p H
\end{equation}
where, M(H) is the magnetization at an applied field H, M$_{\rm S}$ is saturation magnetization, and $\chi_p$ is high field paramagnetic susceptibility. The coefficient of 0.0762 is adopted for cubic anisotropy \cite{Liu06,Herbst98}. The first derivative of the above given equation is:
\begin{equation}
\frac{dM}{dH}=\frac{2(0.0762K_{eff}^2)}{M_SH^3}+\chi_p
\end{equation}					
The $K_{eff}$ was obtained by plotting dM/dH as a function of 1/H$^3$ (not shown) in the high field range for each sample. The variation of $K_{eff}$ with the particle size is shown in Fig.~\ref{Fig5_K_vs_Hc1}(a). We observed a linear relation between $K_{eff}$ and H$_{\rm C}$, as shown in the inset of Fig.~5(a). This confirms the existence of the maxima in $K_{eff}$ and H$_{\rm C}$ at a critical particle size. 

It has been reported that the coercivity increases to a maximum value at a particular particle size and then starts to decrease with further increasing  particle size \cite{Li17,Jiang11}. This is due to a transition from single domain to multi domain regime and the particle size where this transition occurs is called critical size D$_c$. From our data analysis D$_c$ is found in between 12.5~nm and 16~nm. Below D$_c$, the magnetic nanoparticles are in the single domain regime. In this regime the coercivity increases with particle size till D$_c$. At the critical size of the nanoparticles, all the spins are aligned in one direction and this enable a higher magnetic coercivity. In the single domain regime the variation of coercivity with particle size can be written as $H_{\rm C}=g-\frac{h}{D^{3/2}}$, where g and h are constants and D is the particle size \cite{Lee15}. According to this relation the data below D$_c$ were fitted with D$^{-3/2}$, as shown in Fig. \ref{Fig5_K_vs_Hc1}(b). This demonstrated that the nanoparticles with a size from 8.5~nm to about 12.5~nm are in the single domain regime and in this regime the surface effects are more dominating.

With further increasing the particle size the coercivity decreases, which indicates a multi-domain regime above the critical size D$_c$. The particle size variation with coercivity in the multi-domain regime can be expressed as $H_{\rm C}= a+\frac{b}{D}$, where a, b are constants \cite{Lee15}. We have plotted the H$_{\rm C}$ variation with D$^{-1}$ in Fig. \ref{Fig5_K_vs_Hc1}(c) and fitted with this equation, which found to be a straight line. This indicates that above the critical size D$_c$, the observed decrease in H$_{\rm C}$ can be fitted using a multi-domain structure.

\begin{figure}
\includegraphics[width=3.3in]{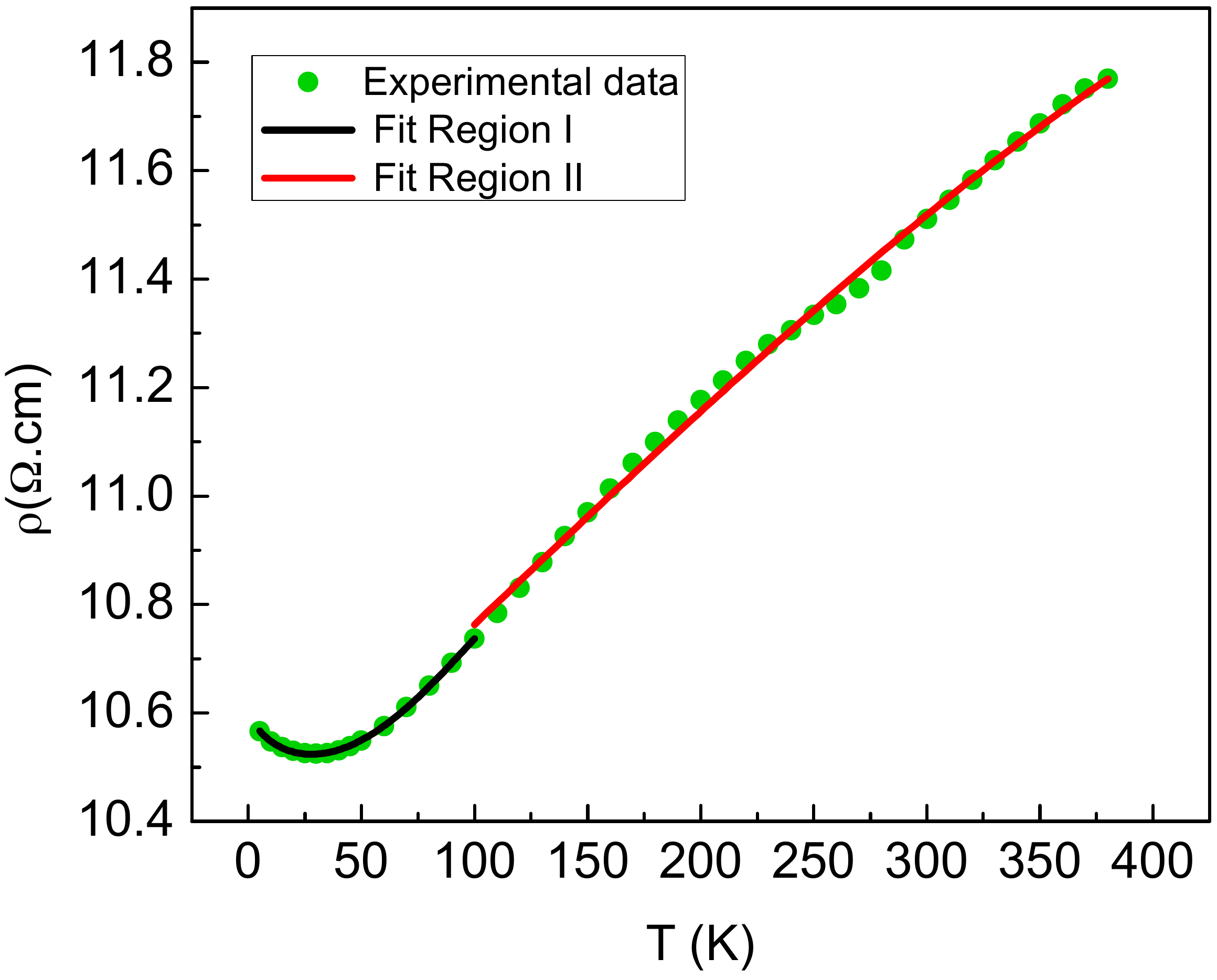}
\caption{Variation of resistivity as a function of temperature for the CFG-25 sample. The solid lines corresponds to the fitted curve in the low (black) and high temperature region (red).}
\label{Fig6_CFG25_RT}
\end{figure}

Here we note that the highest magnetization (closest to the Slater-Pauling rule) is observed for CFG-25 sample. Therefore we have choosen this sample for electrical resistivity, photoemission and Raman measurements. The variation of the resistivity with temperature in 5--380~K range is shown in Fig. \ref{Fig6_CFG25_RT}. The residual resistivity ratio $\rho$$_{300K}$/$\rho$$_{5K}$ is found to be about 1.1, which is lower than the reported value for polycrystalline bulk as well as single crystal Co$_2$FeGa \cite{Zang04, ChenJALCOM08}. The resistivity behavior of single crystal Co$_2$FeGa also shows an upturn at low temperature, but a magnetization value of 5$\mu_B$ is reported \cite{ChenJALCOM08} which is in agreement with Slater-Pauling rule. The mathematical expression for temperature dependent resistivity for a typical ferromagnetic metal is as following \cite{Bainsla14}: 
\begin{equation}
\rho(T) = \rho_0 + \rho_{e-e} + \rho_{e-ph} + \rho_{mag} 
\end{equation}
where $\rho_0$ is the residual resistivity caused by impurities. Electron- electron scattering is expressed by $\rho_{e-e}$, which has a T$^2$ dependence at low temperature. The observed T$^2$ dependence of resistivity could also arise from one magnon scattering, but in half-metals like Heusler alloys, one magnon scattering is absent \cite{Kubo72}. The $\rho_{e-ph}$ from the electron -- phonon interaction has a linear T dependence at high temperatures. The contribution $\rho_{mag}$ arises from either one magnon (T$^2$ dependence) or two magnon (T$^{9/2}$ dependence at low temperatures and T$^{7/2}$ dependence at high temperatures). Another contribution in the resistivity at low temperatures is caused by the disorders in the sample. This contribution can be accounted for the minima followed by an upturn and show a T$^{1/2}$ dependence below 25~K. Therefore, we have analyzed the resistivity in two different regions, i.e. the low temperature region (5~K--100~K) and the high temperature region (100~K--380~K).

\begin{table}[ht]
\caption{The fitted parameter from electrical resistivity data of CFG-25 sample.}
\begin{tabular}{l l l l l l l l} 
\hline\hline 
\multicolumn{4} {l} {Region-I (5~K-100~K)}  \\
\cline{1-4}
A ($\ohm$ cm) & B ($\ohm$ cm K$^{-1/2}$)& C ($\ohm$ cm K$^{-2}$) & D ($\ohm$ cm K$^{-9/2}$) \\
\cline{1-4} 
10.62 & -2.46$\times$10$^{-2}$ & 4.21$\times$10$^{-5}$ & -5.93$\times$10$^{-11}$  \\ 
\hline 
\multicolumn{4}  {l} {Region-II (100~K-380~K)} \\ 
\cline{1-3} 
 & E ($\ohm$ cm)  & F ($\ohm$ cm K$^{-1}$) & G ($\ohm$ cm K$^{-7/2}$) \\
\cline{1-3}
 & 10.35 & 4.05$\times$10$^{-3}$ & -1.21$\times$10$^{-11}$ \\ 
\hline 
\end{tabular}
\label{table2} 
\end{table}

In the temperature range from 5~K to100~K, the resistivity has been fitted with the following relation \cite{Bainsla14, Venkateswarlu16}:
\begin{equation}
\rho(T) = A + B T^{1/2} + C T^{2} + D T^{9/2} 
\end{equation}

For high temperatures between 100~K and 380~K the fitting has been performed using the following equation \cite{Bainsla14, Venkateswarlu16}:
\begin{equation}
\rho(T) = E + F T + G^{7/2}
\end{equation}
Here, the constants A and E represent the residual resistivity value, B is the temperature coefficient for disorder, C and F are the temperature coefficient for $\rho_{e-e}$ and $\rho_{e-ph}$ respectively, D and G are the temperature coefficients for two magnon scattering. All the values of the parameters obtained from the fitting are listed in Table~\ref{table2}. The large values of the coefficients C and F indicate that in the low temperature region the T$^2$ dependence due to disorder dominates, while in the high temperature region, the linear dependence of temperature due to electron phonon interaction dominates. In both the regions there is a weak negative contribution from two magnon scattering.

In order to study the electronic properties of the CFG-25 sample, we have performed x-ray photoemission spectroscopy (XPS) and ultra-violet photoemission spectroscopy (UPS) measurements using a monochromatic AlK$_{\alpha}$ x-ray source (h$\nu=$ 1486.6~eV) and a discharge He source (h$\nu=$ 21.2~eV), respectively. Figures.~7(a--c) show the Co 2$p$, Fe 2$p$ and Ga 2$p$ core level spectra of CFG-25 sample, measured with the same instrumental settings. We have subtracted the inelastic background using the Tougaard method \cite{Tougaard} and then fitted the core level spectra using the Voigt function, which includes instrumental broadenings like Gaussian and Lorentzian contributions from the analyzer, x-ray source, sample temperature, etc. The spin-orbit splitting peaks of Co 2$p$ appear at 778.5~eV and 793.5~eV [see Fig.~7(a)], which corresponds to the 2p$_{3/2}$ and 2p$_{1/2}$ components, respectively. The statistical branching ratio of the spin orbit split peaks i.e. intensity ratio of 2$p_{3/2}$ and 2$p_{1/2}$ is found to be about 2:1, which is close to the expected from the (2$j$+1) multiplicity of the states. In addition, we observed a broad feature at 783~eV i.e. at $\sim$4~eV higher binding energy side of the main Co 2p$_{3/2}$ peak. This broad feature corresponds to the shake-up satellite of Co metal \cite{RaaenSSC86,NathPRB01}. The slight shift in the peak position can be due to the hybridization between Co and Fe states as well as the interactions between core hole and the conduction electrons in Heusler alloys \cite{Alijani11,KozinaPRB14}. On the other hand, Ouardi {\it et al.} explained an observed satellite at about 4~eV higher binding energy than the main Co 2$p_{3/2}$ core level peak caused, which they had observed by hard x-ray photoemission, with exchange splitting \cite{OuardiPRB11}.  

\begin{figure}
\includegraphics[width=3.6in]{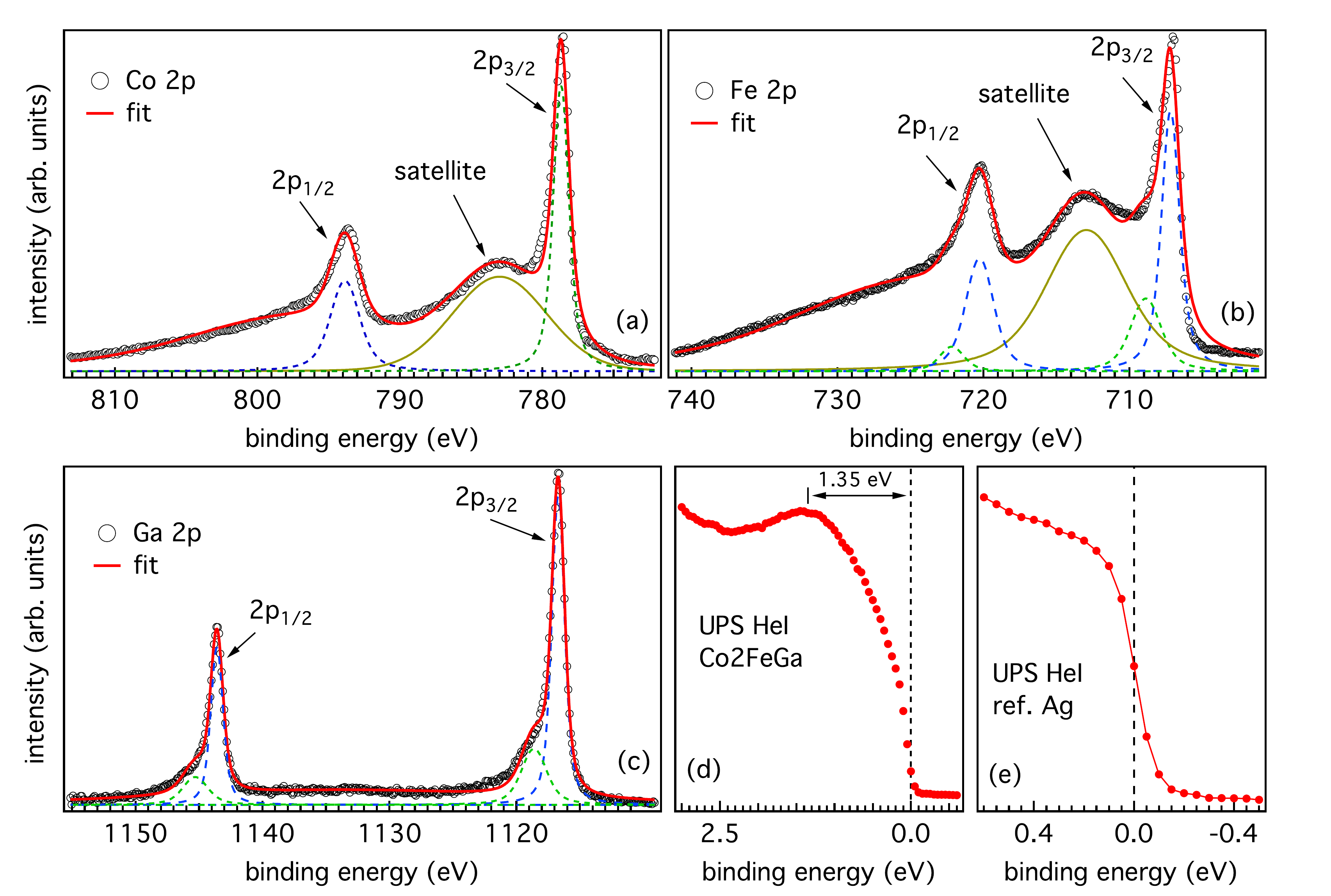}
\caption{The core level XPS spectra (a) Co 2$p$, (b) Fe 2$p$ and (c) Ga 2$p$ of the CFG-25 sample along with the fitted components. (d) UPS valance band spectrum of the CFG-25 sample, and (e) the reference measurement of the Ag Fermi edge measured in same analyzer settings.}
\label{Fig7_XPS_Co2FeGa}
\end{figure}

Furthermore, we observed the peak position of Fe 2$p_{3/2}$ and 2$p_{1/2}$ at about 707~eV and 720~eV binding energies [see Fig.~7(b)], respectively with spin orbit splitting of about 13~eV. Also, the Ga 2$p$ spectrum is shown in Fig.~7(c) with 2$p_{3/2}$ and 2$p_{1/2}$ peaks observed at 1116.7~eV and 1143.5~eV, respectively. These values are at slightly higher binding energies as compared to the elemental Fe 2$p_{3/2}$ (706.8~eV) and Ga 2$p_{3/2}$ (1116.4~eV), respectively. This is possibly due to the hybridization between partially filled 3$d$ levels. We note that a small extra component has been used during the fitting of Fe 2$p$ and Ga 2$p$ core level spectra. The UPS valence band (VB) spectrum of the CFG-25 sample is shown in Fig.~7(d), where the main peak is observed at about 1.5~eV below the Fermi level. The Fermi edge (E$_F$) in valance band spectra of CFG-25 sample has been calibrated with standard Ag sample, mounted with the CFG-25 sample and measured in the same analyzer settings, as shown in Fig.~7(e) for comparison. The spectral shape and peak position of UPS valence band spectrum for CFG-25 sample are in good agreement with other similar Heusler alloys where 3$d$ levels of the transition metals are the dominating contribution \cite{Baral15, Dhaka09, Chakrabarti05, DSouza12}. The main peak in VB spectrum is mainly due to the hybridization between d (e$_g$/t$_{2g}$) sates of Co and Fe. 

\begin{figure}
\includegraphics[width=3.5in]{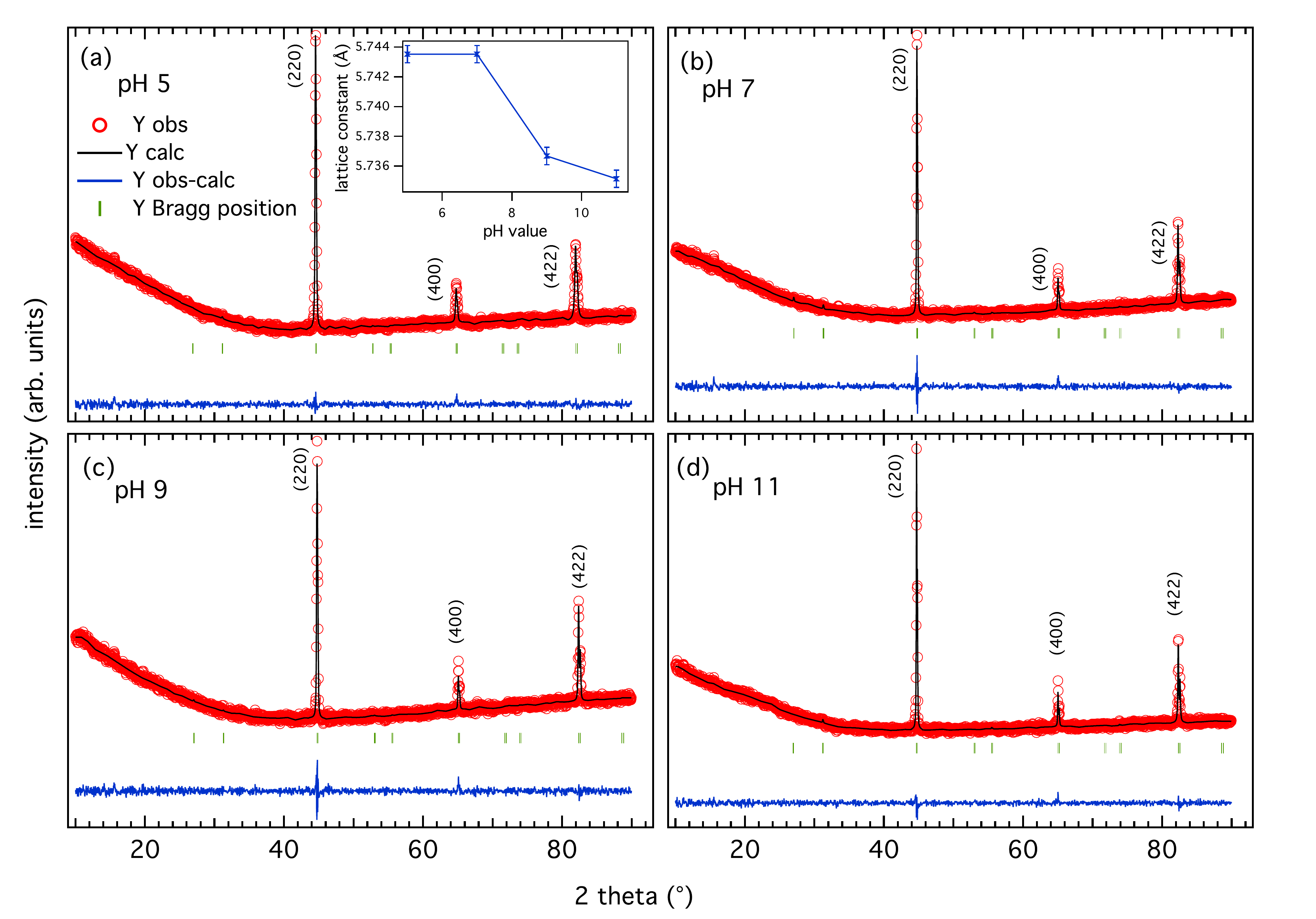}
\caption{(a-d) Powder XRD data (open circles) and Rietveld refinement (black line) of the Co$_2$FeGa samples with different pH values along with the corresponding difference profile (blue line) and Bragg peak positions (vertical green bars). Inset in (a) shows the variation of lattice constant with pH value.}
\label{Fig8_pH_XRD}
\end{figure}

As discussed in the Introduction section, the synthesis process of nanoparticles of Heusler alloys is crucial and the amount of SiO$_2$ in the sol-gel method strongly affects the physical properties. Therefore, we have also prepared Co$_2$FeGa nanoparticles using a co-precipitation method where SiO$_2$ has not required. The XRD patterns for samples prepared with different pH values are shown in Fig. \ref{Fig8_pH_XRD}. From the pattern we can see that all the nanoparticles are well crystalline in nature and the crystal structure is stable having cubic structure without any indication for a secondary phase. The Bragg peaks in XRD pattern corresponds to the (220), (400) and (422) reflections of cubic unit cell and the absence of the (200) and (111) reflections indicates towards the formation of a disorder structure. The Rietveld refined lattice parameter decreases with an increase in pH value as shown in the inset of Fig. \ref{Fig8_pH_XRD}(a). 

\begin{table}[ht]
\caption{Lattice constant ($a$) obtained by Rietveld refinements, saturation magnetizaton (M$_{\rm S}$) and coercivity (H$_{\rm C}$) of Co$_2$FeGa nanoparticles at different pH values.}
\centering
\begin{tabular}{c c c c c} 
\hline\hline 
  & pH5 & pH7 & pH9 & pH11\\ [0.5ex] 
\hline 
$a$ ($\rm \AA$) & 5.7615(5) & 5.7435(4) & 5.7367(4) & 5.7351(3) \\ 
M$_{\rm S}$ ($\mu_{\rm B}/f.u.$) & 3.5 & 4.5 & 5.1 & 6.3\\
$H_{\rm C}$ (Oe) & 21 & 5 & 1 & 14\\
\hline 
\end{tabular}
\label{table3} 
\end{table}

\begin{figure}
\includegraphics[width=3.5in]{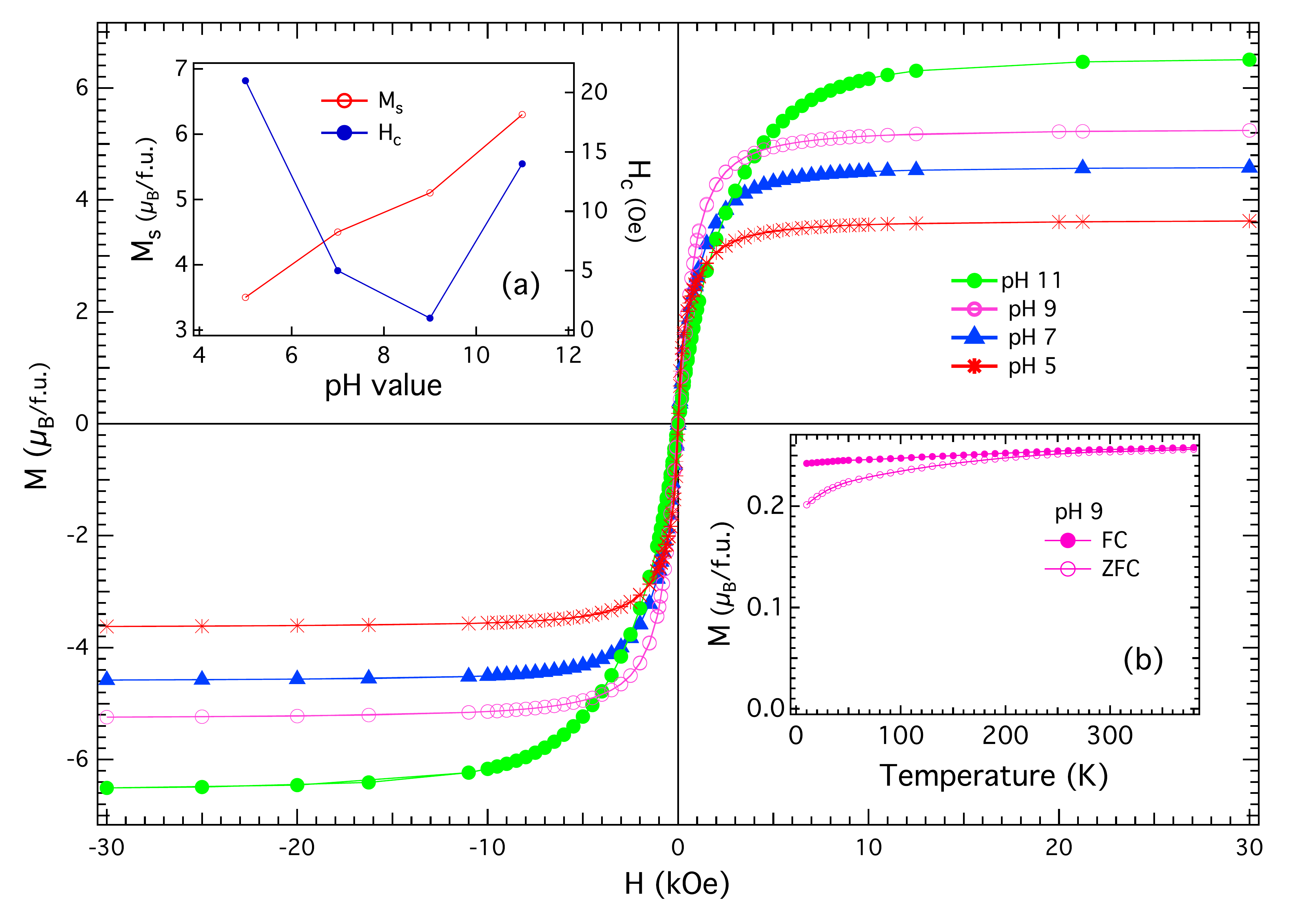}
\caption{Magnetization vs field curves of different samples grown at different pH values measured at 300~K. Inset (a) shows the variation of M$_s$ and H$_c$ with the pH value and (b) shows ZFC and FC curves of the pH 9 sample measured at 100~Oe.}
\label{Fig9_MT_MH_pH}
\end{figure}

In order to study the effect of the pH values on the magnetic properties of the Co$_2$FeGa nanoparticles, temperature and field dependent magnetization measurements have been performed. Fig. \ref{Fig9_MT_MH_pH} shows the H -- M curves of the Co$_2$FeGa nanoparticles prepared at different pH values. The M$_s$ value increases gradually from 3.5 to 6.3 $\mu_{\rm B}/f.u.$ with increasing pH value. This is similar to the previously reported behavior in Al$_3$Fe$_5$O$_{12}$ nanoparticles \cite{Praveena17}. The inset of Fig.~\ref{Fig9_MT_MH_pH} shows the ZFC and FC magnetization curves for nanoparticles synthesized at a pH value of 9, measured between 2~K and 380~K. In both cases (ZFC and FC) the data have been recorded during heating and clearly indicate that the transition temperature is much higher. 

As mentioned before, Co-based Heusler alloys have a large potential for future spintronic applications and the quality of the crystal structure plays an important role in achieving high spin polarization. Therefore, it is also important to study the lattice vibrations, which are highly sensitive to disorder or direct changes in the crystal structure. For this purpose, Raman spectroscopy is a powerful, sensitive and nondestructive technique in order to investigate the structural changes, which provides the information about the modifications in the lattice vibrations. For the Heusler compounds theoretical calculations have predicated that formerly triply degenerate optical modes can split into three well-separated modes: $\Gamma =$ F$_{2g}$ + 2F$_{1u}$ \cite{ZayakPRB05}. Furthermore, out of these modes, only F$_{2g}$ is Raman active and the 2F$_{1u}$ modes are IR active \cite{ZayakPRB05,ManosaPRB01}. On the other hand, there have been several reports that IR-active modes can become Raman active and hence visible in the Raman spectra. Therefore it is possible to see all the optical modes in the Raman spectra of Heusler alloys. However, to the best of our knowledge, there are only two experimental reports on the Raman spectroscopy study on Co-based Heusler alloys \cite{Zhan12,Zhai14}. Zhan {\it et al.}, reported temperature dependent Raman scattering data of Co$_2$Mn$_x$Fe$_{1-x}$Al thin films using the 325~nm laser line as excitation source \cite{Zhan12}. The authors have observed three Raman active modes, which are located at about 320, 569 and 947~cm$^{-1}$. It has been suggested that the vibrational mode at about 320~cm$^{-1}$ corresponds to the F$_{2g}$ phonon. This phonon band can be assigned to the vibrations of the Co atoms in the first sublattice. On the other hand the Raman modes located at 569~cm$^{-1}$ and 947~cm$^{-1}$ can be assigned to the formerly IR-active 2F$_{1u}$ phonons, which involve the vibrations of all atoms together. The authors did not observe change in the intensity and peak position with temperature of the Co-vibration at 320cm$^{-1}$. The absence of a temperature dependence of this phonon mode is unusual and would require further investigations. On the other hand, the high wavenumber Raman modes follow the expected temperature dependence for an anharmonic phonon -- phonon interaction and the expected increase at the magnetic phase transition \cite{Zhan12}. 

\begin{figure}
\includegraphics[width=3.6in]{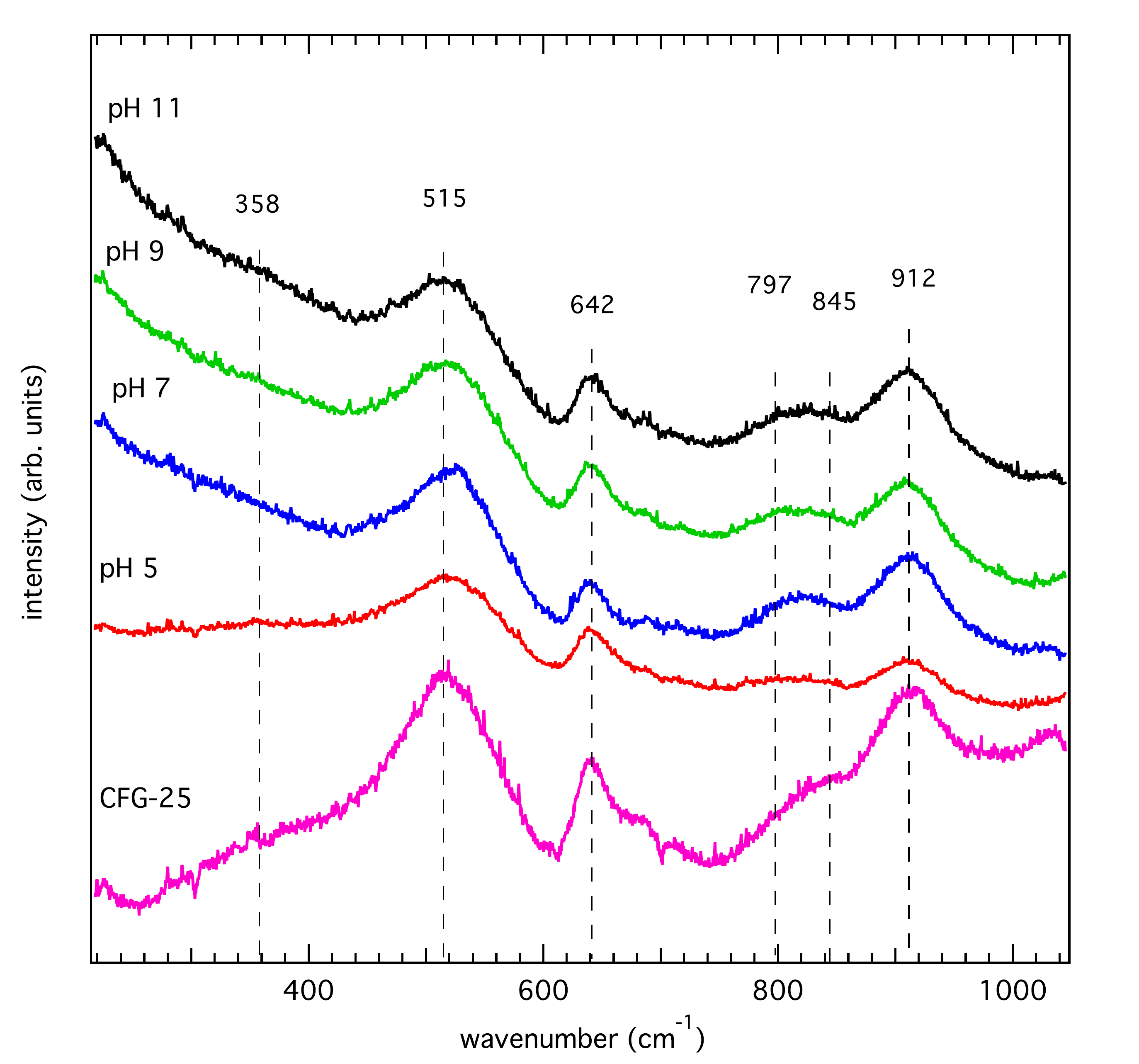}
\caption{The Raman spectra of Co$_2$FeGa nanoparticle samples prepared by sol-gel route (CFG-25) and co-precipitation method (with different pH values, pH5, pH7, pH9, pH11), measured at room temperature using an Ar-ion laser with an excitation wavelength of 514~nm and an excitation power of 10~mW at the sample position.}
\label{Fig10_Raman_allpH}
\end{figure}

Using Raman light scattering we have studied the lattice vibrations of Co$_2$FeGa nanoparticles samples synthesized by chemical route using SiO$_2$ (CFG-25 sample) and different pH values ranging from 5 to 11. In Figure~10 the Raman spectra of these samples are plotted in the wavelength ranges from 200 to 1045~cm$^{-1}$. The measured Raman spectra of all samples are quite similar and six Raman active modes are observed at about 358, 515, 642, 797, 845, and 912~cm$^{-1}$. Recently, Zhai {\it et al.}, observed four Raman active modes in Co$_2$FeAl$_{0.5}$Si$_{0.5}$ films, which are located at 102, 325, 795, and 843~cm$^{-1}$ and are associated with L2$_1$ phase \cite{Zhai14}. Though our XRD results indicate some degree of disorder in the samples, the presence of Raman active modes can be associated with the high chemical ordering of the L2$_1$ structure in all the samples \cite{ZayakPRB05,Zhan12}. However, the presence of the lower symmetry chemical structures with B2 and A2 types cannot be fully ruled out since these phases do not possess Raman-active modes. It must be noted that all phonon modes are quite broad. Also, the first principles investigation of phonon softening of Heusler alloys predicts three phonon modes for the L2$_1$ crystal structure \cite{ZayakPRB05}. On the other hand, we observed six modes and also Zhai {\it et al.} observed five modes \cite{Zhai14}, which indicate that the formerly IR active modes become Raman active. These can be caused by either chemical disorder in the samples or by effects due to the nanoscale size of the particles, i.e. enhanced surface effects, broken translational symmetry, or non-uniform internal strain. By comparing with refs.~\cite{Zhan12,Zhai14}, our data suggest that the observed Raman peaks correspond to T$_{1u}$/F$_{1u}$ modes, which are generally found in Heusler alloys \cite{ZayakPRB05,ManosaPRB01} and are related to the vibrations of all the atoms together. The difference in the peak positions between the bulk and nanoparticle samples are a direct indication for the effects of the nanoscale size of the particles. Further experiments are required to precisely determine the crystal structure of the nanoparticles where new functionalities such as anisotropies in the spin polarization for the different crystallographic orientations might emerge. Our room temperature Raman data on metallic Heusler alloys should serve as a further motivation to study the behavior of phonon modes.

\section{\noindent ~Conclusions}

In conclusion, we have successfully synthesized ternary Co$_2$FeGa nanoparticles via two different chemical methods. First, we optimized the synthesis procedure of these nanoparticles by sol-gel method using different SiO$_2$ concentrations and have analyzed their structural, magnetic, transport and electronic properties. The XRD, SAED, and high resolution TEM data confirm the perfect crystalline single phase of all nanoparticles. The particle size of these spherical nanoparticles was controlled by varying the SiO$_2$ concentration. We have established the correlation between particle size and domain structure of these nanoparticles, where the coercivity H$_{\rm C}$ increases with increasing particle size, reaches a maximum value at a critical size, and then decreases with further increase in particle size. The effective magnetic anisotropy constant also behaves in this manner. These results are discussed in terms of the transition from single domain to multi-domain regime. The contribution to electrical resistivity in CFG-25 sample is mainly caused by disorder and electron -- phonon interactions in the low and high temperature region, respectively. The valance band spectrum of CFG-25 sample shows a hybridization between the $d$ states of Co and Fe. The Co$_2$FeGa nanoparticles prepared by co-precipitation method show magnetization increases with increase in pH value. Finally we have observed Raman active modes, which can be attributed to the high chemical ordering L2$_1$ structure in all our samples. On the other hand, the L2$_1$ symmetry can be broken in the nanoparticles due to the size effects, internal strain and enhanced surface effects.

\section{\noindent ~Acknowledgments}

PN acknowledges the MHRD, India for fellowship. Authors thank IIT Delhi for various experimental facilities: XRD, PPMS EVERCOOL-II at Physics department, TEM and HRTEM at Central Research Facility (CRF). We also thank the physics department of IIT Delhi for excellent support. RSD gratefully acknowledges the financial support from BRNS through DAE Young Scientist Research Award project sanction No. 34/20/12/2015/BRNS. RSD also grateful to the department of science \& technology (DST), India for support through Indo-Australia early and mid-career researchers (EMCR) fellowship, administered by INSA (sanction order no. IA/INDO-AUST/F-19/2017/1887) and UNSW for hosting the visit. CU acknowledges the support of the Australian Research Council (ARC) through the Discovery Grant DP160100545. We thank Yousef Kareri for help during Raman and photoemission measurements .

\end{document}